\pdfoutput=1
\newif\ifFull
\Fullfalse
\documentclass[letterpaper, 11 pt, onecolumn, conference]{ieeeconf}  
\IEEEoverridecommandlockouts                              

\usepackage{graphicx}
\usepackage{url}
\usepackage{times}
\usepackage{epstopdf}
\usepackage{subfigure}
\usepackage{amsmath}                                                           
\usepackage{amssymb}                                                           
\ifFull
\usepackage{algorithmic}
\else
\usepackage[noend]{algorithmic}
\fi

\makeatletter
\def\@begintheorem#1#2{\sl \trivlist \item[\hskip \labelsep{\bf #1\ #2:}]}
\def\@opargbegintheorem#1#2#3{\sl \trivlist
      \item[\hskip \labelsep{\bf #1\ #2\ #3:}]}
\makeatother

\newcommand{\E}{{\bf E}}

\newcommand{\eat}[1]{}

\begin{document}
\title{\LARGE \bf Analyzing Distributed Join-Idle-Queue: \\ 
A Fluid Limit Approach}
\author{Michael Mitzenmacher{$^1$}\thanks{{$^1$}School~of Engineering and Applied Sciences, Harvard University.  Supported in part by the NSF under grants CCF-1535795, CCF-1320231, and CNS-1228598.}}
\date{}
\maketitle
\thispagestyle{empty}
\pagestyle{empty}

\begin{abstract}
In the context of load balancing, Lu et al. \cite{Luetal} introduced
the distributed Join-Idle-Queue algorithm, where a group of 
dispatchers distribute jobs to a cluster of parallel servers.  Each
dispatcher maintains a queue of idle servers; when a job arrives to a
dispatcher, it sends it to a server on its queue, or to a random
server if the queue is empty.  In turn, when a server has no jobs, it
requests to be placed on the idle queue of a randomly chosen
dispatcher.

Although this algorithm was shown to be quite effective, the original
asymptotic analysis in \cite{Luetal} makes simplifying assumptions
that become increasingly inaccurate as the system load increases.
Further, the analysis does not naturally generalize to interesting
variations, such as having a server request to be placed on the idle queue
of a dispatcher before it has completed all jobs, which can be beneficial under high
loads.

We provide a new asymptotic analysis of Join-Idle-Queue systems based
on mean field fluid limit methods, deriving families of differential
equations that describe these systems.  Our analysis avoids previous
simplifying assumptions, is empirically more accurate, and generalizes naturally to the variation 
described above, as well as other simple variations.  Our theoretical
and empirical analyses shed further light on the performance of Join-Idle-Queue,
including potential performance pitfalls under high load.  
\end{abstract}

\section{Introduction}
\label{sec:intro}
The Join-Idle-Queue (JIQ) algorithm is an approach designed to
effectively approximate Join-the-Shortest-Queue (JSQ) algorithm in
large-scale distributed systems.  In this setting jobs coming into a
system are sent to one of a group of {\em dispatchers}; the
dispatchers are in turn responsible for sending the job to a {\em
server} within a cluster of parallel servers for processing.  For
the Join-the-Shortest-Queue algorithm, each of the dispatchers would
have to keep up-to-date load information on all of the servers, which
could be undesirable because of the required communication.  For the
Join-Idle-Queue algorithm, a server informs a dispatcher when it is
idle, having completed all of its jobs.  The dispatchers each keep a queue
of idle servers, dubbed {\em I-queues} to distinguish them from queues
for jobs.  When a job arrives, it sends the job to an idle server from
its I-queue, or to a server chosen uniformly at random if its I-queue is
empty.  Further details on the model are given in Section~\ref{sec:model}.

The goal of JIQ is to provide excellent load balancing -- most jobs
are immediately sent to an idle queue -- while providing scalability
for systems of thousands of servers or larger.  In particular, the
communication requirements of JIQ are low, and importantly the task of
determining the load of various servers is not on the critical path
when assigning a job.  As noted in \cite{Luetal}, alternative schemes
requiring a job to communicate to determine the load of servers on the
arrival of the job can increase the overall response time, and hence 
may not be suitable for many practical systems.  On the theoretical side, the two-layer approach of
load balancing idle servers among dispatchers to allow jobs to be load
balanced among servers provides interesting challenges for analysis.

JIQ systems were introduced in \cite{Luetal}, and analyzed in the large system
limit where the number of servers grows to infinity and the ratio
between the number of servers and the number of dispatchers is kept fixed.  
The analysis presented there, however, assumes that there are only
idle processors in the dispatchers' I-queues.  This assumption is
inaccurate; indeed, as noted in \cite{Luetal}, it ``is violated when
an idle server receives a random arrival.''  They argue that idle
servers receive such arrivals relatively rarely, and show empirically
that their approximation appears accurate (under sufficiently low
arrival rates).

Here we provide a more direct mean-field fluid limit approach that
provides an accurate analysis of the large system limit without the
assumptions of the previous analysis by deriving a family of
differential equations that corresponds to the behavior large system
limit.  Besides yielding a proper analysis of the basic JIQ system
analyzed in \cite{Luetal}, this approach allows straightforward
analysis of many generalizations.  For example, under high loads, it
may be better to have a server be placed on an I-queue when it has one
job left (or, more generally, $k$ jobs left), rather than waiting for
the server to become idle, since it may take some time for the I-queue
to present the server with a new job.  In \cite{Luetal} it is left as
an open question to analyze such variations.  We generalize our
equations to this setting here, as well as other variations, such as
using last-come first-served instead of first-come first-served at the
I-queues in order to determine which server to send a job.  (In
contrast to the analysis of \cite{Luetal}, we find once simplifying
assumptions are removed, such choices can make a small but noticeable
difference.)

Finally, in both our theory and our empirical work, we examine some
potential pitfalls of JIQ systems.  In particular, under high load,
there can be insufficiently many idle queues available, causing
degradations in performance as the dispatchers distribute jobs
randomly.  Even when the expected waiting time remains smaller than
alternatives, JIQ systems can exhibit significantly higher variance,
which may be unsuitable for some workloads.  As we show later in the
paper, having a server place itself on an I-queue when it has one job
left greatly mitigates this problem, and we would recommend this variation
when there is any possibility that the system may enter a highly loaded
state.  

The main limitation of our analysis is that it applies to the setting
of exponentially distributed service times and a Poisson arrival
process.  While generalizations to other service times and arrival
processes are theoretically possible by representing them as mixtures
of exponentials (see e.g. \cite{mitzenmacher2001power,MitzThesis}),
or by considering more general partial differential equations
(see e.g. \cite{mfd,stolyar2015pull,asymp-vved})
a more general analysis of JIQ for more general service and arrival 
processes remains an interesting problem.  

Also, we emphasize that here our goal is to derive the appropriate
equations, examine implications for the use of JIQ systems and their
variations, and demonstrate empirically the accuracy of this approach.
One can also formally prove convergence mathematically, but this is
not our main goal.  There appears to be no impediment to utilizing
standard techniques for mean field fluid limit analysis of queueing
systems (as in \cite{MitzThesis,ShwartzWeiss,stolyar2015pull,q-vved})
for JIQ systems, so we leave this aspect to the inclined reader.


\section{Preliminaries:  Model and Notation}
\label{sec:model}

Here we follow much of the model and notation of \cite{Luetal}.  We
have a system with $n$ servers and $m$ dispatchers (or equivalently
I-queues), where we fix the ratio $r = n/m$ and let $n,
m \rightarrow \infty$.  Jobs arrive as a Poisson process of total rate
$n\lambda$, where $\lambda < 1$, and are sent to a dispatcher chosen
uniformly at random, so at each dispatcher jobs arrive as a Poisson
process of rate $n\lambda/m$.  The service times are exponentially
distributed with mean 1.  Each dispatcher has an I-queue, which is an
ordered list of servers.  If the I-queue is non-empty, the dispatcher
sends the arriving job to a server on the I-queue; we start by
considering the I-queues being governed by the first-come first-served
(FCFS) discipline.  If the I-queue is empty, the dispatcher sends the
job to a server chosen uniformly at random.  A server is placed on an
I-queue when it finishes all of its jobs (so its job queue is empty)
and it is not already on an I-queue, so that at any  time it is 
at most on only a single I-queue.  We are assuming here that the dispatcher notes when
sending a job whether it was randomly assigned from an empty I-queue,
so that a server can track when it is removed from an I-queue, 
but such an implementation is straightforward.

There are multiple ways in which a server can choose which I-queue to
join.  In \cite{Luetal}, two natural variations are studied.  First,
on each completion a server can join an I-queue chosen uniformly at random;
this is referred to as JIQ-Random.  
Alternatively, a server can probe the length of $d$ I-queues and choose
the one with the smallest number of jobs.  This approach of using a constant
number of choices for placement
 has been studied in many settings,
in particular in the context where jobs choose one of a collection
of parallel servers for service on arrival, and is referred to as SQ$(d)$.  (The approach is
sometimes referred to as the power of two choices, or balanced
allocations \cite{MitzThesis,mitzenmacher2001power,TwoSurvey}.  The variation where jobs arrive to parallel servers and
choose one according to SQ$(d)$ is often referred to as the
``supermarket model'' \cite{MitzThesis}.)  In the context of JIQ, this variation
is referred to as JIQ-SQ$(d)$.  Using SQ$(d)$ to distribute idle servers
among I-queues helps ensure that I-queues do not empty, avoiding the
need to send incoming jobs to random servers instead of idle ones.

\section{The JIQ-Random Model}

\subsection{Derivation of Limiting Equations for the JIQ-Random Model}
\label{sec:JIQR}

In the equations that follow, we consider the evolution of I-queues and
servers.  Recall $m$ is the numbers of I-queues, $n$ the number of servers,
and $r=n/m$.  As done in \cite{Luetal}, we consider the regime where $r$ is held constant and
$n$ and $m$ are growing to infinity.  We use $Q_i(t)$ to represent the
number of I-queues with $i$ queued servers, and $q_i(t) =
Q_i(t)/m$ be the corresponding fraction of I-queues with $i$ queued servers.
For $i \geq 0$ and $j \geq 1$, we use $S_{i,j}(t)$ to
represent the number of servers with $i$ jobs that are in position $j$ in
one of the I-queues, and similarly we use $s_{i,j}(t)$ to represent the
corresponding fraction of servers.  Finally, we use
$S_{i,\varnothing}(t)$ to represent the number of servers with $i$
jobs that are not currently in an I-queue, and similarly we
use $s_{i,\varnothing}(t)$ to represent the corresponding fraction of
servers.  

To begin, consider a short time interval $dt$, and 
consider the change in $Q_i$ over that time interval.  We use 
$\Delta Q_i(t)$ to denote $\E[Q_i(t+dt)-Q_i(t)]$.  Then for suitably short
time intervals, we take the probability of an arrival at the I-queues to be $\lambda n dt$
and the total probability of a service event at servers to be $n dt$.  Then for $i\geq 1$,
\begin{eqnarray*}
\Delta Q_i(t) & = & \frac{Q_{i+1}(t)-Q_i(t)}{m} \lambda n dt - \frac{Q_{i}(t)-Q_{i-1}(t)}{m}S_{1,\varnothing}(t) dt.
\end{eqnarray*}
This equation represents that $Q_i$ increases when an arrival occurs at a I-queue with $i+1$ servers,
and correspondingly decreases when an arrival occurs at a I-queue with $i$ servers.
Similarly, $Q_i$ increases when a server not already queued completes its last job and is placed
on a I-queue with $i-1$ servers,
and correspondingly decreases when a server not already queued completes its last job and is placed
on a I-queue with $i-1$ servers.
Dividing through by $dt$ and using $r=n/m$ yields
\begin{eqnarray*}
\frac{\Delta Q_i(t)}{dt} & = & \lambda r (Q_{i+1}(t)-Q_i(t)) - r(Q_{i}(t)-Q_{i-1}(t))s_{1,\varnothing}(t).
\end{eqnarray*}
Dividing through by $m$, and considering the fluid limit, where we replace the expected change over time
with the corresponding differential equation, we find
\begin{eqnarray}
\label{eqn:qi1}
\frac{dq_i}{dt} & = & \lambda r (q_{i+1}(t)-q_i(t)) - r(q_{i}(t)-q_{i-1}(t))s_{1,\varnothing}(t).
\end{eqnarray}

For $i=0$ we have the corresponding equation 
\begin{eqnarray}
\label{eqn:q01}
\frac{dq_0}{dt} & = & \lambda r q_{1}(t) - rq_{0}s_{1,\varnothing}(t).
\end{eqnarray}
Note that there is no $-\lambda r q_0(t)$ term, since an arrival at an empty I-queue
is immediately dispatched to a random server, leaving the I-queue empty.  

We now consider the $s_{i,j}$.  We use $\Delta S_{i,j}(t)$ to denote
$\E[S_{i,j}(t+dt)-S_{i,j}(t)]$, where $j$ can include the value
$\varnothing$.  For $i \geq 1$ and $j \geq 1$, we find
\begin{eqnarray*}
\Delta S_{i,j}(t) & = & (S_{i+1,j}(t)-S_{i,j}(t))dt + (\lambda n dt)q_0(t)(s_{i-1,j}(t)-s_{i,j}(t)) + (\lambda n dt)\frac{S_{i,j+1}(t)-S_{i,j}(t)}{m}.
\end{eqnarray*}
The first term, $(S_{i+1,j}(t)-S_{i,j}(t))dt$, represents the change in $S_{i,j}(t)$ due to the completion
of jobs at the servers, changing the number of jobs at the server.  The second term is the probability
that a new job arrives to an empty I-queue, and then is sent randomly to a server with either $i-1$ or $i$ jobs that
is $j$th on its I-queue.  The last term represents the change in $S_{i,j}(t)$  due to an arrival that reduces
the I-queue length at one of the I-queues holding a server that contributes to $S_{i,j}(t)$ or $S_{i,j+1}(t)$.  
This equation simplifies in the fluid limit to
\begin{eqnarray}
\label{eqn:sij1}
\frac{ds_{i,j}(t)}{dt} & = & s_{i+1,j}(t)-s_{i,j}(t) + \lambda q_0(t)(s_{i-1,j}(t)-s_{i,j}(t))
+ \lambda r (s_{i,j+1}(t)-s_{i,j}(t)).  
\end{eqnarray}

The case where the number of jobs at the server is 0 is similar:
\begin{eqnarray}
\label{eqn:s0j1}
\frac{ds_{0,j}(t)}{dt} & = & s_{1,j}(t) - \lambda q_0(t)s_{0,j}(t) + s_{1,\varnothing}(t)q_{j-1}(t)
+ \lambda r (s_{0,j+1}(t)-s_{0,j}(t)).  
\end{eqnarray}
Here the term $s_{1,\varnothing}q_{j-1}$ corresponds to servers with one job and not on an I-queue 
finishing their job and then joining an I-queue with $j-1$ other servers. 

Finally, we also need to handle servers that are enqueued.
For $i\geq 2$,
\begin{eqnarray}
\label{eqn:si01}
\frac{ds_{i,\varnothing}(t)}{dt} & = & s_{i+1,\varnothing}(t)  - s_{i,\varnothing}(t)
+ \lambda q_0(t) (s_{i-1,\varnothing}(t)-s_{i,\varnothing}(t)) + \lambda r s_{i-1,1}(t).  
\end{eqnarray}
Here the first terms $s_{i+1,\varnothing}(t)  - s_{i,\varnothing}(t)$ correspond to job
completions at servers with $i+1$ or $i$ jobs that are not on an I-queue;  the next
terms $\lambda q_0(t) (s_{i-1,\varnothing}(t)-s_{i,\varnothing}(t))$  correspond to an
arrival at an empty I-queue that lands at a server not on an I-queue;  and the 
term $\lambda r s_{i-1,1}(t)$ corresponds to an arrival at an I-queue where the job
at the front goes from having $i-1$ jobs to $i$ jobs.    

Similarly, for $i = 1$,
\begin{eqnarray}
\label{eqn:s101}
\frac{ds_{1,\varnothing}(t)}{dt} & = & s_{2,\varnothing}(t)  - s_{1,\varnothing}(t)
- \lambda q_0(t) s_{1,\varnothing}(t) + \lambda r s_{0,1}(t).  
\end{eqnarray}

\subsection{Equilibrium in the Limiting JIQ-Random Model}

An equilibrium distribution for the above system of equations
satisfies that all the derivatives equal 0.  We show how to find the
equilibrium distribution; let us use ${\bar s}_{i,j}$ and ${\bar q}_i$
for the corresponding values in equilibrium.
Our process will be to show that all values ${\bar s}_{i,j}$ can 
be written in terms of ${\bar s}_{1,\varnothing}$, and then that there is 
a unique value of ${\bar s}_{1,\varnothing}$ that can be found numerically
that satisfies equilibrium conditions.

Various equilibrium conditions arise from first principles and from the
equations above.  
For example, in an equilibrium, $\sum_{j \geq 1} {\bar s}_{0,j} = 1-\lambda$, 
since the fraction of idle queues must be $1-\lambda$.

Equations~(\ref{eqn:qi1}) and~(\ref{eqn:q01}) yield 
$${\bar q}_{i+1} = {\bar q}_i \frac{{\bar s}_{1,\varnothing}}{\lambda};$$ in
an equilibrium ${\bar s}_{1,\varnothing} < \lambda$, since $\lambda$ is the
total fraction of servers that are not idle.  It follows that
${\bar s}_{1,\varnothing} = \lambda(1-{\bar q}_0)$.  

The final key equation for determining the equilibrium distribution is the following:
for $i \geq 0$ and $j \geq 1$, 
$${\bar s}_{i,j+1} = {\bar s}_{i,j} \frac{{\bar s}_{1,\varnothing}}{\lambda}.$$ 
To see this, consider a server with $i$ jobs and $j$th in queue at some time $t$.  The last time it chose an I-queue, 
suppose instead it had chosen an I-queue with length one larger than the one it actually chose, and then consider 
coupling all subsequent events except for interchanging the events (arrivals, servers joining) at these two I-queues.
From this coupling, and from the fact that the ratio between the equilibrium probability of consecutive I-queue lengths
is constant, we see that the relative probability that this server is $j$th in its I-queue instead of $(j+1)$st
is in fact the ratio between the equilibrium probability of consecutive queue lengths, 
${\bar s}_{1,\varnothing}/{\lambda}$.  

We then have
\begin{eqnarray*}
1-\lambda & = & \sum_{j \geq 1} {\bar s}_{0,j} \\
          & = & \sum_{k \geq 0} {\bar s}_{0,1} \left (\frac{{\bar s}_{1,\varnothing}}{\lambda} \right )^k\\
          & = & {\bar s}_{0,1} \frac{\lambda}{\lambda-{\bar s}_{1,\varnothing}}.
\end{eqnarray*}
Hence 
$${\bar s}_{0,1} = \frac{(\lambda-{\bar s}_{1,\varnothing})(1-\lambda)}{\lambda},$$
and 
$${\bar s}_{0,j} = \frac{(\lambda-{\bar s}_{1,\varnothing})(1-\lambda){\bar s}_{1,\varnothing}^{j-1}}{\lambda^{j}}.$$

From Equations~(\ref{eqn:s101}) and~(\ref{eqn:si01}), we can now solve for the values ${\bar s}_{i,\varnothing}$.  
Setting the derivative to 0 in Equation~(\ref{eqn:s101}) yields 
\begin{eqnarray*}
{\bar s}_{2,\varnothing} & = & {\bar s}_{1,\varnothing} + \lambda {\bar q}_0 {\bar s}_{1,\varnothing} - \lambda r {\bar s}_{0,1} \\  
& = & {\bar s}_{1,\varnothing} (1 + \lambda {\bar q}_0) - r(\lambda-{\bar s}_{1,\varnothing})(1-\lambda) \\
& = & {\bar s}_{1,\varnothing} (1 +r(1-\lambda) + \lambda-{\bar s}_{1,\varnothing}) - r\lambda(1-\lambda).
\end{eqnarray*}
Similarly, setting the derivative to 0 in Equation~(\ref{eqn:si01}) yields 
\begin{eqnarray*}
{\bar s}_{i+1,\varnothing} & = & {\bar s}_{i,\varnothing}
- \lambda {\bar q}_0 ({\bar s}_{i-1,\varnothing}-{\bar s}_{i,\varnothing}) - \lambda r {\bar s}_{i-1,1},  
\end{eqnarray*}
and inductively we can solve for all ${\bar s}_{i+1,\varnothing}$ in terms of 
${\bar s}_{1,\varnothing}$.  

Equation~(\ref{eqn:s0j1}) allows us to find ${\bar s}_{1,1}$ in terms of ${\bar s}_{1,\varnothing}$  
by looking at when $j=1$ and setting
the derivative to 0:
\begin{eqnarray*}
{\bar s}_{1,1} & = & \lambda {\bar q}_0(t){\bar s}_{0,1} - {\bar s}_{1,\varnothing}{\bar q}_{0}
- \lambda r ({\bar s}_{0,2}-{\bar s}_{0,1}).  
\end{eqnarray*}
Equation~(\ref{eqn:s0j1}) then again allows us to find the remaining values of ${\bar s}_{i,j}$ inductively,
in terms of ${\bar s}_{1,\varnothing}$.  

We have tested these equations by doing a binary search to find the
appropriate value of ${\bar s}_{1,\varnothing}$ that leads to a
solution where the sum of the ${\bar s}_{i,j}$ values equals one, and
have found it matches the results found in our empirical evaluation
below in Section~\ref{sec:verify}.  We have not, to this point, been
able to prove formally that there is unique fixed point equilibrium to
these equations, although our testing of these equations thus far
suggests that this is the case.  We have also not yet found a way to
express the equilibrium in a convenient form in terms of $\lambda$.

\subsection{Empirical Verification of the Equations for the JIQ-Random Model}
\label{sec:verify}

To test the accuracy of our equations, we ran a simulation of the
actual queueing system.  In our results below (and throughout the
paper), we started at an empty
state and ran for 10000 units of time, with 10000 servers and 1000
dispatchers, so that $r = 10$.  (We chose $r=10$ as this was the
primary setting used in simulations in \cite{Luetal}.)  We focus on the average time in
system for a job for convenience, although our simulations show more
generally that our equations also track the load distribution of
servers and I-queues accurately as well.  We report the average time
for simulations for jobs that complete after time 5000 and before time
10000; this gives the system time to reach approximate
equilibrium, for suitable comparison.  In reporting the average time in system, we take the
average of 1000 trials; we note the standard deviation over experiments is small.  The
results from our family of differential equations are derived from the
average server load and Little's Law.  We started the differential equations
with all I-queues empty and all servers with one job initially for convenience.
We used the Euler method with a step size of $0.01$;  experiments with other
step sizes led to slightly different results for the average time in system 
but we found the differences were $0.01\%$ or less; we therefore felt
the step size was appropriate for presenting results.

Finally, we also compare with the results from \cite{Luetal}.  Specifically, for the case of
exponentially distributed service times and Poisson arrivals, based
on their assumptions they derive an analytical expression for the
average time in system, given by  
\begin{eqnarray}
\label{luform}
1 + \frac{\lambda}{(1-\lambda)(1+r)}.
\end{eqnarray}

Our results of Table~\ref{tab:table1} demonstrate that our fluid limit approach
is significantly more accurate, particularly under higher loads, as
one might expect given the assumptions use in \cite{Luetal}.  In these
experiments, the magnitude of our relative error is, in the worst
example, about $0.1\%$, while for the previous approximation the
magnitude of the relative error ranges from roughly $1$ to $6\%$.

\begin{table}
\begin{center}
\begin{tabular}{|c|c|c|c|}
\hline
$\lambda$ & Sims &  Diff &      Formula \\
          &      &   Equations &  (\ref{luform}) \\ \hline
0.50 & 1.12886 & 1.12894 &  1.09091 \\  \hline
0.60 & 1.17987 & 1.17995 &  1.13636 \\  \hline
0.70 & 1.25888 & 1.25895 &  1.21212 \\  \hline
0.80 & 1.40787 & 1.40790 &  1.36364 \\  \hline
0.90 & 1.83712 & 1.83659 &  1.81818 \\  \hline
0.95 & 2.68138 & 2.68035 &  2.72727 \\  \hline
0.96 & 3.10314 & 3.10086 &  3.18182 \\  \hline
0.97 & 3.80509 & 3.80110 &  3.93939 \\  \hline
0.98 & 5.20555 & 5.20069 &  5.45455 \\  \hline
0.99 & 9.40217 & 9.39754 &  10.0000 \\  \hline
\end{tabular}
\caption{The differential equation method provides accurate estimates of the average time
in system for JIQ-Random at 10000 queues, with relative error of about 0.1\% at $\lambda = 0.99$
and smaller relative error for lower rates.  The results are much more accurate than
the still quite good approximation derived in \cite{Luetal}.}
\label{tab:table1}
\end{center}
\end{table}

We have also compared the results from the differential equations with
the results we obtain by calculating the equilibrium distribution, and
find that they are in agreement to several decimal places; e.g.,
$s_{1,0}(t)$ and ${\bar s}_{1,0}$ agree for $t=$10000.  The equilibrium
distribution calculation does suffer from numerical precision challenges
for small values of ${\bar s}_{i,j}$, but we leave this issue for further work. 

\section{Variations of the JIQ Model} 

In this section, we consider several variations on JIQ that can be
modeled using differential equations in a manner similar to our
model for JIQ-Random presented in Section~\ref{sec:JIQR}.
Specifically, we consider having servers place themselves on I-queues
before becoming idle, using JIQ-SQ$(d)$, and utilizing other
disciplines besides FCFS for the I-queues.  One can also naturally
combine these variations; here we consider them separately to
simplify the exposition.  Our focus here is in showing how to
generate the relevant equations; we leave additional questions (such
as the nature of the equilibrium distribution) for future work.

\subsection{Early Assignment of Servers to I-Queues}

We may modify the JIQ-Random model to handle servers placing
themselves on I-queues before they are idle, such as when they have
just one job remaining.  Such an approach appears potentially
beneficial under very high loads when there are (on average) fewer
idle servers than I-queues, since empty I-queues default to the poor
strategy of assigning a job to a processor chosen uniformly at random.
To make the intuition more concrete, note that when $r=10$ and $\lambda >
0.9$, on average less than $1/10$ of the servers will be idle, leading
to empty I-queues.

Below we modify the equations for the
setting where any time a server is not on an I-queue, and its load
falls to a threshold $z$ or below, it places itself on an I-queue.
We refer to this as the {\em early threshold} variation.
While one could also consider a richer framework for I-queues, which
either sorted incoming servers by their current load or tracked the
load of servers placed on them, here we provide the equations for the
simplest generalization.

Deriving a method to analyze this type of strategy was specifically
left as an open question in \cite{Luetal}.  They examined one such
strategy empirically, where their strategy allowed a server to be
placed on multiple I-queues.  Analyzing strategies where servers can
be on multiple I-queues would necessarily complicate our analysis, as then
we need to use variables that track the position of the servers in multiple
I-queues (e.g., ``three-dimensional'' variables instead of the ``two-dimensional''
$s_{i,j}$).  Although such analysis is possible, we consider here only the
natural generalization where a server can only place itself on one I-queue
at a time.

The equations for the I-queues remain nearly the same, but now the queue length
increases whenever a server not on an I-queue completes servicing a job and has $z$ or fewer jobs
remaining.  
\begin{eqnarray*}
\frac{dq_i}{dt} & = & \lambda r (q_{i+1}(t)-q_i(t)) - r(q_{i}(t)-q_{i-1}(t))\sum_{k=1}^{z+1}s_{k,\varnothing}(t).
\end{eqnarray*}

\begin{eqnarray*}
\frac{dq_0}{dt} & = & \lambda r q_{1}(t) - rq_{0}\sum_{k=1}^{z+1}s_{k,\varnothing}(t)
\end{eqnarray*}

The equations relating to $s_{i,j}$ do not change for $i > z$;  but for $i \leq z$, we must 
consider that a server with $i$ jobs not on an I-queue will be placed on an I-queue on a job
completion.    Hence for $i > z$, $j \geq 1$, we have  
\begin{eqnarray*}
\frac{ds_{i,j}(t)}{dt} & = & s_{i+1,j}(t)-s_{i,j}(t) +  \lambda q_0(t)(s_{i-1,j}(t) - s_{i,j}(t))
+ \lambda r (s_{i,j+1}(t)-s_{i,j}(t)).  
\end{eqnarray*}
And for $1 \leq i \leq z$, $j \geq 1$, we have 
\begin{eqnarray*}
\frac{ds_{i,j}(t)}{dt} & = & s_{i+1,j}(t) -s_{i,j}(t) + \lambda q_0(t)(s_{i-1,j}(t) - s_{i,j}(t))
+ s_{i+1,\varnothing}(t)q_{j-1}(t) + \lambda r (s_{i,j+1}(t)-s_{i,j}(t)).  
\end{eqnarray*}
Finally, for $i = 0$, $j \geq 1$ we have 
\begin{eqnarray*}
\frac{ds_{0,j}(t)}{dt} & = & s_{1,j}(t) -\lambda q_0(t) s_{0,j}(t) + s_{1,\varnothing}(t)q_{j-1}(t)
+ \lambda r (s_{0,j+1}(t)-s_{0,j}(t)).  
\end{eqnarray*}

For servers not on a I-queue, the equations remain the same for $i >
z$.  For $i \leq z$, however, we must now account for the fact that a
server with $i$ jobs not on an I-queue will be placed on an I-queue on
a job completion.  Hence for $i > z$, $j =\varnothing$ we have
\begin{eqnarray*}
\frac{ds_{i,\varnothing}(t)}{dt} & = & s_{i+1,\varnothing}(t)  - s_{i,\varnothing}(t)
+ \lambda q_0(t) (s_{i-1,\varnothing}(t) - s_{i,\varnothing}(t)) + \lambda r s_{i-1,1}.  
\end{eqnarray*}
For $1 \leq i \leq z$, $j =\varnothing$ we have
\begin{eqnarray*}
\frac{ds_{i,\varnothing}(t)}{dt} & = & -s_{i,\varnothing}(t)
+ \lambda q_0(t) (s_{i-1,\varnothing}(t) - s_{i,\varnothing}(t)) + \lambda r s_{i-1,1}.  
\end{eqnarray*}

In Table~\ref{tab:table2} we examine this variation for the setting $z=1$,
again comparing simulation results to the differential equations, as well as
to the simulation results for JIQ-Random.  The relative error is slightly higher,
but the results are again extremely accurate.
The key insight, beyond again the 
accuracy of the fluid limit approach, is that under high load performance
improves dramatically by allowing non-idle servers to queue at I-queues.  
Of course this advantage would be less important for a larger $r$.  We have
also tried $z=2$, and obtained slightly better performance than $z=1$ at $\lambda = 0.99$,
but worse performance at lower values of $\lambda$.

\begin{table}
\begin{center}
\begin{tabular}{|c|c|c|c|}
\hline
$\lambda$ & Sims &  Diff &      JIQ- \\
          & $z=1$  &   Equations &  Random \\  \hline
0.50 & 1.19356 & 1.19369 &  1.12886 \\  \hline
0.60 & 1.28239 & 1.28254 &  1.17987 \\  \hline
0.70 & 1.40813 & 1.40828 &  1.25888 \\  \hline
0.80 & 1.59780 & 1.59796 &  1.40787 \\  \hline
0.90 & 1.94387 & 1.94377 &  1.83712 \\  \hline
0.95 & 2.33613 & 2.33539 &  2.68138 \\  \hline
0.96 & 2.48165 & 2.48041 &  3.10314 \\  \hline
0.97 & 2.69113 & 2.68903 &  3.80509 \\  \hline
0.98 & 3.04536 & 3.04146 &  5.20555 \\  \hline
0.99 & 3.91664 & 3.90627 &  9.40217 \\  \hline
\end{tabular}
\caption{Comparing the average time in system for JIQ-Random with an early threshold of 1 with results
from the differential equations and JIQ-Random.  The differential
equations yield accurate results, and under high rates performance
improves significantly.}
\label{tab:table2}
\end{center}
\end{table}

\subsection{Varying the I-Queue Discipline}

Intuitively, we might think that first-come first-served might not be the best 
queuing discipline for the I-Queues, at least under higher loads.  This is because
servers that have been waiting in I-Queues may have received randomly assigned jobs
and may no longer be idle;  queues that have just arrived to an I-queue may be more
likely to be idle.

Our fluid limit approach allows us to write equations for at least
some other I-queue disciplines.  Here we analyze 
last-come first-served (LCFS), with a slight change to our equations and interpretations of the
variables.  (We can similarly analyze the process where I-queues choose a server in their
queue uniformly at random for an arriving job, for example.) 
Here for $i \geq 0$ and $j \geq 1$, we use $S_{i,j}(t)$ to
represent the number of servers $j$th from the front of an 
I-queue and with $i$ jobs;  with LCFS when a server joins an I-queue
it is placed at the front of the I-queue and all other jobs move one step further
from the front.  We use $s_{i,j}(t)$ to represent the
corresponding fraction of servers.  We use
$S_{i,\varnothing}(t)$ to represent the number of servers with $i$
jobs that are not currently in an I-queue, and 
use $s_{i,\varnothing}(t)$ to represent the corresponding fraction.  
The analysis is similar to that in Section~\ref{sec:JIQR}, so we simply provide the final equations.  

\begin{eqnarray*}
\frac{dq_i}{dt} & = & \lambda r (q_{i+1}(t)-q_i(t)) - r(q_{i}(t)-q_{i-1}(t))s_{1,\varnothing}(t).
\end{eqnarray*}
\begin{eqnarray*}
\frac{dq_0}{dt} & = & \lambda r q_{1}(t) - rq_{0}s_{1,\varnothing}(t).
\end{eqnarray*}
\eat{
For $i \geq 2$ and $j \geq 1$, 
\begin{eqnarray*}
\Delta S_{i,j}(t) & = & (S_{i+1,j}(t)-S_{i,j}(t))dt + (\lambda n dt)q_0(t)(s_{i-1,j}(t)-s_{i,j}(t)) + (\lambda n dt)\frac{S_{i,j+1}(t)-S_{i,j}(t)}{m} + S_{1,\varnothing}(t)\frac{S_{i,j-1}(t)-S_{i,j}(t)}{m}.
\end{eqnarray*}
This equation simplifies in the fluid limit to
}
For $i \geq 1$ and $j \geq 2$, 
\begin{eqnarray*}
\frac{ds_{i,j}(t)}{dt} & = & s_{i+1,j}(t)-s_{i,j}(t) + \lambda q_0(t)(s_{i-1,j}(t)-s_{i,j}(t))
+ \lambda r (s_{i,j+1}(t)-s_{i,j}(t)) + r s_{1,\varnothing}(t)(s_{i,j-1}(t)-s_{i,j}(t)).  
\end{eqnarray*}

For $i \geq 1$ and $j = 1$, 
\begin{eqnarray*}
\frac{ds_{i,1}(t)}{dt} & = & s_{i+1,1}(t)-s_{i,1}(t) + \lambda q_0(t)(s_{i-1,1}(t)-s_{i,1}(t))
+ \lambda r (s_{i,2}(t)-s_{i,1}(t)) - r s_{1,\varnothing}(t)s_{i,1}(t).  
\end{eqnarray*}

\eat{The case where the number of jobs at the server is 0 is similar:}
For $i = 0$ and $j \geq 2$, 
\begin{eqnarray*}
\frac{ds_{0,j}(t)}{dt} & = & s_{1,j}(t) - \lambda q_0(t)s_{0,j}(t) + rs_{1,\varnothing}(t)s_{0,j-1}
- rs_{1,\varnothing}(t)s_{0,j}
+ \lambda r (s_{0,j+1}(t)-s_{0,j}(t)).  
\end{eqnarray*}

For $i = 0$ and $j = 1$, 
\begin{eqnarray*}
\frac{ds_{0,1}(t)}{dt} & = & s_{1,1}(t) - \lambda q_0(t)s_{0,1}(t) 
+ \lambda r (s_{0,2}(t)-s_{0,1}(t))
- rs_{1,\varnothing}(t)s_{0,1} + s_{1,\varnothing}.
\end{eqnarray*}

For $i\geq 2$,
\begin{eqnarray*}
\frac{ds_{i,\varnothing}(t)}{dt} & = & s_{i+1,\varnothing}(t)  - s_{i,\varnothing}(t)
+ \lambda q_0(t) (s_{i-1,\varnothing}(t)-s_{i,\varnothing}(t)) + \lambda r s_{i-1,1}.  
\end{eqnarray*}

For $i = 1$,
\begin{eqnarray*}
\frac{ds_{1,\varnothing}(t)}{dt} & = & s_{2,\varnothing}(t)  - s_{1,\varnothing}(t)
- \lambda q_0(t) s_{1,\varnothing}(t) + \lambda r s_{0,1}.  
\end{eqnarray*}

Table~\ref{tab:table3} demonstrates that there is a noticeable if small
performance difference between LCFS and FCFS when using JIQ-Random, with
the intuition that LCFS would perform better being correct.  Moreover,
the differential equations are able to accurately capture this small
performance gap.  

\begin{table}
\begin{center}
\begin{tabular}{|c|c|c|c|}
\hline
$\lambda$ & Sims &  Diff &      JIQ- \\
          &      &   Equations &  Random \\ 
          & (LCFS)  &  (LCFS) & (FCFS) \\  \hline
0.50 & 1.10976 & 1.10980 &  1.12886 \\  \hline
0.60 & 1.15732 & 1.15379 &  1.17987 \\  \hline
0.70 & 1.23305 & 1.23310 &  1.25888 \\  \hline
0.80 & 1.37805 & 1.37796 &  1.40787 \\  \hline
0.90 & 1.79941 & 1.79893 &  1.83712 \\  \hline
0.95 & 2.63751 & 2.63559 &  2.68138 \\  \hline
0.96 & 3.05663 & 3.05429 &  3.10314 \\  \hline
0.97 & 3.75659 & 3.75259 &  3.80509 \\  \hline
0.98 & 5.15449 & 5.15006 &  5.20555 \\  \hline
0.99 & 9.35407 & 9.34465 &  9.40217 \\  \hline
\end{tabular}
\caption{Comparing average time in system for JIQ-Random with LCFS against FCFS I-queues.  
The differential equations capture the performance difference between
the two approaches.}
\label{tab:table3}
\end{center}
\end{table}

\subsection{JIQ-SQ$(d)$}
Recall that for JIQ-SQ$(d)$, a server chooses which I-queue to join by 
choosing $d$ I-queues at random and joining the least loaded (breaking ties randomly).    
The changes required to model the use of SQ$(d)$ for I-queues are minimal.  
However, it becomes easier utilizing a standard change of notation, to
keep track of the {\em tails} of the distribution \cite{MitzThesis,q-vved}.  That is, let $\hat{q}_i(t)$
be the fraction of I-queues with at least $i$ jobs, instead of exactly $i$ jobs,
at time $t$.  Note $\hat{q}_0(t)=1$.

To begin, to see the change of notation, for the JIQ-Random model, we have for $i\geq 1$:
\begin{eqnarray*}
\frac{d{\hat{q}}_i}{dt} & = & -\lambda r ({\hat{q}}_i(t) - {\hat{q}}_{i+1}(t)) +  r({\hat{q}}_{i-1}(t)- {\hat{q}}_{i}(t)) s_{1,\varnothing}(t).
\end{eqnarray*}
For the JIQ model using SQ$(d)$ for I-queues, all that changes from the I-queue standpoint is when a server finishes, the
probability a server lands on a I-queue with $i-1$ other jobs becomes
$(({\hat{q}}_{i-1}(t))^d- ({\hat{q}}_{i}(t))^d)$.  Hence for the generalization we have
\begin{eqnarray*}
\frac{d{\hat{q}}_i}{dt} & = & -\lambda r ({\hat{q}}_i(t) - {\hat{q}}_{i+1}(t)) +  r(({\hat{q}}_{i-1}(t))^d- ({\hat{q}}_{i}(t))^d) s_{1,\varnothing}(t).
\end{eqnarray*}
Correspondingly, from the server side, the only change is to the equation for $\frac{ds_{0,j}(t)}{dt}$
\begin{eqnarray*}
\frac{ds_{0,j}(t)}{dt} & = & s_{1,j}(t) - \lambda q_0(t)s_{0,j}(t) + s_{1,\varnothing}(t)(({\hat{q}}_{j-1}(t))^d - ({\hat{q}}_{j}(t))^d)
+ \lambda r (s_{0,j+1}(t)-s_{0,j}(t)).  
\end{eqnarray*}
Notice that for convenience we have used the expression $q_0(t)$ in the equations, although one could write all equations
in terms of the ${\hat{q}}_i(t)$ as needed.  

The results of Table~\ref{tab:table4} will be perhaps at this point
unsurprising.  The differential equations prove quite accurate, and
the performance JIQ-SQ$(d)$ is better than JIQ-Random. However,
under high loads the improvements are arguably small because the
key issue that there are not enough idle servers to fill the I-queues
remains.  Distributing the small number of idle servers better does not
address this issue as well as the early threshold approach, which increases
the number of servers being placed on I-queues.

\begin{table}
\begin{center}
\begin{tabular}{|c|c|c|c|}
\hline
$\lambda$ & Sims &  Diff &      JIQ- \\
          & JIQ-SQ$(2)$ &   Equations &  Random \\  \hline
0.50 & 1.01029 & 1.01033 &  1.12886 \\  \hline
0.60 & 1.02377 & 1.02379 &  1.17987 \\  \hline
0.70 & 1.05558 & 1.05557 &  1.25888 \\  \hline
0.80 & 1.14044 & 1.14027 &  1.40787 \\  \hline
0.90 & 1.46106 & 1.46035 &  1.83712 \\  \hline
0.95 & 2.19241 & 2.19045 &  2.68138 \\  \hline
0.96 & 2.57696 & 2.57420 &  3.10314 \\  \hline
0.97 & 3.23186 & 3.22894 &  3.80509 \\  \hline
0.98 & 4.57810 & 4.57186 &  5.20555 \\  \hline
0.99 & 8.71553 & 8.70009 &  9.40217 \\  \hline
\end{tabular}
\caption{Comparing average time in system for JIQ-SQ$(2)$ and JIQ-Random.  The differential
equations yield accurate results, and there is an improvement 
over JIQ-Random, notably under lower loads.}
\label{tab:table4}
\end{center}
\end{table}

\section{Variance in JIQ Systems}

In \cite{Luetal}, the authors compare JIQ systems with supermarket
model, or SQ$(d)$, systems, where there are no I-queues; instead, each
incoming job probes the load at $d$ randomly chosen servers, and is
assigned to the least loaded server from its $d$ choices.  They note
that JIQ systems often perform better, in terms of average time in
system.  This is true when the system load is sufficiently low, as
most jobs are assigned to idle servers.  However, under very high
load, when I-servers are often empty, we see that the default to
random assignment can significantly increase average time in system,
so that it becomes worse than SQ$(d)$.  (This phenomenon motivated our
examination of the early threshold variation.)

Even when JIQ systems perform better in terms of average time in
system, however, the variance of JIQ systems can be higher than
comparable SQ$(d)$ systems.  In particular there are likely to be some
number of ``unlucky'' servers that receive a large number of randomly
assigned jobs, and thereby find themselves with long queues.  For
example, in all of the JIQ systems, an arrival is assigned randomly
with probability $q_0(t)$, and hence if ${\bar q}_0$ represents an
equilibrium probability of an empty I-queue, servers obtain randomly
assigned arrivals at a rate of $\lambda {\bar q}_0$.  As such, in JIQ
systems the tails of queues fall at most geometrically, while in
SQ$(d)$ systems they are known to fall doubly
exponentially \cite{MitzThesis,q-vved};  for example, even with
a supermarket model with just two choices, the fraction of queues with load $j$ or greater
falls like $\lambda^{2^{j-1}}$, in the asymptotic limit as the system grows large.
The maximum loaded server is therefore likely to have significantly more jobs in JIQ
systems under high loads.  

These results are apparent in both the results of the fluid limit models
and in simulations.  Here we present simulation results for convenience
in Table~\ref{tab:table5}, although the results are entirely similar
using the results from the differential equations to obtain the variance
for an arriving job at time 10000 (when the system is essentially at equilibrium).    
We compare SQ$(2)$ with JIQ-Random, and with JIQ-Random with a threshold
of one.  The high variance of JIQ-Random is apparent, as well as its poor
relative performance under very high loads.  The early threshold greatly
mitigates these flaws with JIQ, although under sufficiently high loads
even with the early threshold the variance increases beyond SQ$(2)$, as
expected.   

\begin{table}
\begin{center}
\begin{tabular}{|c|c|c|c||c|c|c|}
\hline
$\lambda$ & SQ$(2)$ & JIQ    &   JIQ- &  SQ$(2)$ & JIQ    &   JIQ-  \\
         &          & $z=1$  &  Random   &          & $z=1$  &  Random \\  
         &  Avg     & Avg  &  Avg          &  Var     & Var  &  Var \\  \hline
0.50 & 1.26572 & 1.19356 & 1.12886 & 1.49234 & 1.39094 & 1.26742 \\  \hline
0.60 & 1.40747 & 1.28239 & 1.17987 & 1.74541 & 1.56715 & 1.37965 \\  \hline
0.70 & 1.61446 & 1.40813 & 1.25888 & 2.11398 & 1.81205 & 1.56210 \\  \hline
0.80 & 1.94744 & 1.59780 & 1.40787 & 2.70517 & 2.17632 & 1.93743 \\  \hline
0.90 & 2.61442 & 1.94387 & 1.83712 & 3.86808 & 2.86939 & 3.26202 \\  \hline
0.95 & 3.38375 & 2.33613 & 2.68138 & 5.14687 & 3.81179 & 6.94056 \\  \hline
0.96 & 3.64960 & 2.48165 & 3.10314 & 5.56989 & 4.22673 & 9.31200 \\  \hline
0.97 & 4.00222 & 2.69113 & 3.80509 & 6.11533 & 4.89627 & 14.0488 \\  \hline
0.98 & 4.51796 & 3.04536 & 5.20555 & 6.88176 & 6.23157 & 26.4497 \\  \hline
0.99 & 5.43592 & 3.91664 & 9.40217 & 8.16057 & 10.6235 & 87.1031 \\  \hline
\end{tabular}
\caption{Comparing JIQ-Random and JIQ-Random with an early threshold of 1 with 
SQ$(d)$, for both the average and the variance of the time in system.  The early threshold reduces
both the average and variance under high load, although under high enough
loads even the early threshold version has higher variance than SQ$(d)$.}
\label{tab:table5}
\end{center}
\end{table}

Table~\ref{tab:table5} is not the entire story.  When we look at the
maximum load over our simulation, we also see the performance difference
clearly, even with an early threshold.  For SQ$(2)$, the maximum load
at the end time $10000$ over all of our simulations when $\lambda =
0.99$ is 10, and when $\lambda = 0.8$ it is 6.  For JIQ-Random with
an early threshold $z = 1$, the maximum load at the end time of our
simulation when $\lambda = 0.99$ is always over 20, and is sometimes
over 30; when $\lambda = 0.8$, it is 11.

We emphasize that under larger values of $r$, which may be suitable for
real-world systems, these concerns are lessened, as higher arrival rates
are needed for there to be large numbers of empty I-queues.  

\section{Conclusion} 

We have derived fluid limit equations that describe the
distributed Join-Idle-Queue process under Poisson arrivals and exponentially
distributed service times, showing that these systems are amenable to
asymptotic analysis.  Our methodology allows us to obtain very
accurate estimates for reasonably sized finite systems.  By examining
variations, we have shown the generality of this method, and in
particular we have analyzed the compelling approach of serving
I-queues while still processing one or more jobs.  Based on its reasonable
performance under low loads and its significantly better performance in terms
of both expectation and variance under
higher loads, we recommend strongly that builders of systems that want to use the
distributed JIQ approach examine the use of an early threshold.  

On the theoretical side, for many of the JIQ variations we have examined, it is worthwhile to consider whether
the asymptotic equilibrium distributions have a useful or pleasant form
that could aid in rapidly calculating system performance or making comparisons
among system.  We leave this for future work.  

\bibliographystyle{plain}

\end{document}